\documentclass[useAMS,usenatbib,dcolumn,usegraphicx,twocolumn,10pt]{mnarxiv}
\usepackage{amsmath,amssymb,mathtools}
\usepackage[dvips]{graphicx}
\usepackage{xspace}
\usepackage{soul}
\usepackage{ulem}
\usepackage{float}
\newcommand{\abs}[1]{\left|#1\right|}
\newcommand{\br}[1]{\left(#1\right)}
\renewcommand{\eqref}[1]{Eq.\ref{#1}}
\newcommand{\figref}[1]{Fig.\ref{#1}}
\newcommand{\grad}{\frac{\km}{\mathrm{s}\cdot\,\kpc}}
\newcommand{\HI}{H{\scriptsize I}\xspace}
\newcommand{\km}{\mathrm{km}}
\newcommand{\kpc}{\mathrm{kpc}}
\newcommand{\mpc}{\mathrm{Mpc}}

\newcommand{\ngc}{NGC 4244\xspace}
\renewcommand{\sq}[1]{\left[#1\right]}

\newcommand{\ud}[1]{\mathrm{d}#1}

\title[]{Modeling vertical structure in circular velocity of~spiral~galaxy~NGC~4244}
\author[]{
{Joanna Ja{\l}ocha,$^{1}$}
{{\L}ukasz Bratek,$^{1}$\thanks{corresponding author: Lukasz.Bratek@ifj.edu.pl}}
{Szymon Sikora,$^{2}$}
{Marek Kutschera$^{3}$}
\\
$^{1}$Institute of Nuclear Physics,
Polish Academy of Sciences, Radzikowskego 152, PL-31342 Krak\'{o}w, Poland\\
$^{2}$Astronomical Observatory, Jagiellonian University, Orla 171, PL-30244 Krak\'{o}w, Poland\\
$^{3}$Institute of
Physics, Jagellonian University,  Reymonta 4, PL-30059 Krak{\'o}w, Poland}

\begin{document}

\date{\today}
\pagerange{\pageref{firstpage}--\pageref{lastpage}} \pubyear{}

\maketitle

\begin{abstract}
We study the vertical gradient in azimuthal velocity of
spiral galaxy NGC 4244 in a thin disk model. With surface
density accounting for the rotation curve,
 we model the gradient properties in the approximation of quasi-circular orbits and find the predictions to be consistent with the gradient properties inferred from measurements.
This consistency may suggest that the mass distribution in this galaxy is flattened.
\end{abstract}



\section{Introduction}

Measurements of galactic kinematics carries the information of how
gravitating mass is distributed in galactic interiors. A basic quantity encoding this information is a rotation curve. But for a spiral galaxy appropriately aligned with
respect to the observer, it is also possible to ascertain and study the vertical structure of the rotation above the mid-plane which provides a further piece of information
indispensable for a more precise determination of the galactic  mass
distribution.

Measurements enabling a more detailed detection of the velocity field structure
have been performed, among others, for spiral galaxies: NGC 891
\citep{bib:Oosterloo,bib:Benjamin}, NGC 4559
\citep{bib:Barbieri4559}, NGC 4302 \citep{bib:Heald4302}, NGC 5775
\citep{bib:ngc5775}, for NGC 2403 \citep{bib:fraternali2403} and
for the Milky Way galaxy \citep{bib:gradmlecz}. In particular, it has been established  that the azimuthal component of the rotation
gradually diminishes in almost a linear fashion with the altitude above the mid-plane.

Recently, we tested the hypothesis that
spiral galaxies, in contrast to
what seems to be true for most galaxies, do not necessarily
have to be dominated by massive non-baryonic dark matter halos -- at
least it cannot be excluded that some of these galaxies might be flattened disk-like
objects. But, in the absence of dominating spherical components, the motion of matter in the galactic mid-plane vicinity, to within the approximation of axial symmetry, should be satisfactorily well  described as if
governed entirely by the gravitational field of a thin (or a finite-width) axi-symmetric disk with surface (or column) density
accounting for the observed fragment of the rotation curve. From a good mass model it is additionally required that its predictions should be consistent with other observations, in particular, the model should predict correct behavior in the structure of the galactic rotation field. In this context we may use the vertical gradient observable as a test of the disk model of spiral galaxies.

We dealt with the vertical gradient in
disk model in \citep{bib:gradient} and
\citep{bib:gradientwie}.
To describe the gradient in this model we additionally assumed that
the projections of orbits onto the mid-plane were approximately circular,
which is the subject of the so called {\it quasi-circular orbits
approximation}  assumed also in the present paper.
In \citep{bib:gradient} we also verified our approach
by performing a numerical simulation of motion
of test bodies on various orbits in the gravitational potential  of a thin disk and we
compared the analytical estimates of our approach with  the simulated fall-off in azimuthal velocity near the
mid-plane. The
results of the simulation and of our approach were consistent with
each other.
In the thin disk model framework we were able to reproduce
the observed values of the gradient and its properties, such as a linear decrease of  azimuthal velocity with the altitude above the mid-plane. We consider this as a success of the thin disk model.

Furthermore, in \citep{bib:jalocha2014} we applied a more general,
finite-width disk model to study the vertical gradient in the
Milky Way. The width of the disk was constrained so as to obtain
the best fit to microlensing measurements being another source of information about the mass distribution, quite independent of the rotation curve. On this occasion we had the opportunity to test limits on
the applicability of the thin disk to modeling the vertical gradient of rotation. The conclusion was intuitively clear: when the gradient
measurements were performed not too close to the mid-plane, outside the main concentration of mass of the finite-width disk (for
heights at least $0.5\,\kpc$ in the case of the Milky Way galaxy),
then the approximation of the
infinitesimally thin disk was sufficient and there was no need for using
the computationally more demanding finite-width disk.

The two observations above are the basis for our use of the thin
disk modeling of the vertical gradient in the present paper. We expect that the approach could be applicable also to other
flattened galaxies.

\medskip

\ngc is another
galaxy for which the vertical structure in the rotational velocity can be studied.
\citet{bib:ngc4244} published a
velocity field above the mid-plane as well as the resulting rotation
curve in both galactic halves.
The reported
results for the detected gradient in their numerical model of \ngc were comparable in both galactic halves: $-9^{+3}_{-2}\grad $ in the
approaching half and $-9^{+2}_{-2} \grad $ in the receding half, with the
gradient decreasing in magnitude to $-5^{+2}_{-2}\grad $ and $-4^{+2}_{-2} \grad $ near a radius
of $10\,\kpc$ in these halves, respectively.
The authors pointed
out that the lag in \HI was present in absence of an \HI halo.
This is an indication that the mass distribution in \ngc may be flattened, and is what motivates our applying of the thin disk to model this particular galaxy.
We
want to check whether our simple model applied to \ngc
would account for the reported gradient values
and their variability with the radial distance.

\section{Mass distribution in thin disk}

In disk model
we are looking for a surface
density accounting for the entire observed rotation curve under the assumption of axial and mid-plane-reflection symmetries.

For obvious reasons, rotation curves do not extend far enough and
one has to extrapolate the rotation data. But then,
the model results unavoidably depend on the way one chooses to extrapolate beyond the last measured point. This
feature is not a failure of a mass model as such but is
inherent in the gravitation of flattened mass distributions. In general, for a flattened mass distribution (if not representable as a union of contiguous similar homeoids), the  rotation value at a given radius is depended also on the masses
distributed exterior to that radius, likewise, column mass density at a
given point is a point-dependent integral functional of the entire rotation
curve. Therefore, the thin disk, as a model of a flattened mass distribution, must be used with due care. A more detailed discussion of these
issues were described by \citet{bib:bratek_MNRAS}. Sometimes, it is
possible to find a global solution by iterations when additional
measurements complementary to the rotation data  are known beyond the last
measurement of the rotation \citep{bib:jalocha_apj}.

\subsection{Rotation curve of \ngc}

As for \ngc, we take the rotation curve  as the
starting point. We make use of the rotation measurements
published by \citet{bib:ngc4244}. They were carried out
separately for both halves of the galaxy, but we have superimposed and averaged out
these data because the modeling should be done consistently with
the axial symmetry the disk model
approximation postulates. This approximation should suffice to model
the vertical gradient to the approximation of axial symmetry, because the gradient values inferred from measurements, as it was mentioned
earlier, are comparable in both halves.

\begin{figure}
   \centering
      \includegraphics[width=0.5\textwidth
      ,trim=16pt 18pt 4pt 16pt
      ,clip=true
      ]{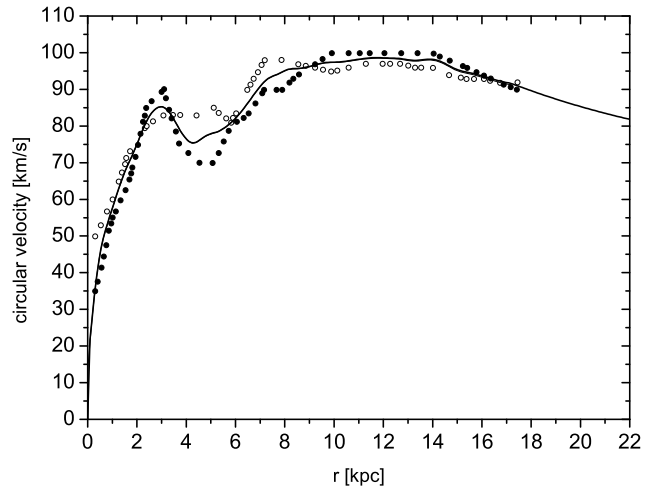}
      \caption{\label{fig:rot}
Circular velocity of \ngc in the mid-plane as a function
of the galactocentric distance (obtained by assuming a distance to \ngc of $4.4\,\mpc$). The collection of points show the rotation values in the approaching {\it [black circles]}
and receding {\it [open circles]} halves read from curves published in \citep{bib:ngc4244}.
The {\it [solid line]} is our adopted rotation curve obtained by means of  interpolating between the rotation values in both halves and extended beyond $\approx17\,\kpc$ by adjoining a Keplerian tail.}
  \end{figure}

The rotation curve which we adopt in this work is shown in figure  \figref{fig:rot}.
It is obtained by interpolating the measured rotation values and, additionally, by extending them beyond $\approx17\,\kpc$ by a Keplerian tail. The purpose of this artificial extension is only to facilitate
integration to infinity in finding the corresponding surface density, and it will not significantly affect the
surface density at lower radii in the region of interest for the gradient estimation. This extension is not entirely arbitrary, however. A decrease in a
Keplerian fashion in the rotation, in a neighborhood of the last
measured point, was reported in another work investigating \ngc \citep{1996AJ....112..457O}. This observation may be used to support
the form of the artificial extension we adopted.

\subsection{Surface density in \ngc}

In modeling the mass distribution and the vertical gradient, we limit ourselves to using the thin disk model,  instead of applying a finite-width
disk, because, as it was illustrated by \citet{bib:jalocha2014},
when the gradient is calculated above the mid-plane, outside the
main concentration of masses characterized by some vertical
width-scale, then the vertical structure becomes insignificant for
the gradient determination in this region and one can use a
simpler model.

We determine a surface density in thin disk model
based on a given circular velocity, by using a formula
 turning the velocity squared to the
surface density as derived by \citet{2014arXiv1411.0197B}:
\begin{equation}  \label{eq:SigmafromRotCrv}\!
\sigma(\rho)=\frac{1}{2\pi^2 G}\int\limits_{0}^{\infty}\sq{ \frac{K\!\br{\frac{2\sqrt{x}}{1+x}}}{1+x}+
\frac{E\!\br{\frac{2\sqrt{x}}{1+x}}}{1-x}}
\frac{v^2( x^{-1}\rho)}{\rho}\,\ud{x}.
\end{equation}
This integral should be evaluated in the principal value sense
about the singularity $x=1$ of the expression in the square brackets.\footnote{We use complete elliptic integrals $K$
and $E$ in forms as defined by \citet{eliptic}: \[
K(k)=\int\limits_{0}^{\pi/2}\frac{\ud{\phi}}{
\sqrt{1-k^2\sin^2{\phi}}},\qquad
E(k)=\int\limits_{0}^{\pi/2}\ud{\phi} \sqrt{1-k^2\sin^2{\phi}}.\]
Note, that they differ in the adopted convention for the argument notation from those given by \citet{1972hmfw.book.....A}.}

As we have already pointed this out, the surface mass density is uncertain close to the
radial position of the last point on the rotation curve and beyond it, owing to to the nature of a flattened mass distribution as such and  lack of measurement data. In some cases we could overcome this indeterminacy by additionally
using a surface density of gas distributed beyond the last point and
find a global rotation by iterations in a way similar to that proposed by \citet{bib:jalocha_apj}.
But there is no such complementary measurements available for
\ngc, and we have to choose a different approach.

Since we are
interested in the gradient modeling in the galactic interior $r<10\,\kpc$ separated well enough  from the
last measured point at $\approx17\,\kpc$, we chose to extrapolate the rotation by smoothly joining to it
a Keplerian tail, so as to extend the natural slope of the curve
to larger radii, as shown in \figref{fig:rot}. The internal part of the corresponding surface density is not significantly dependent on this extension to within reasonable distortions of that extension. To support this statement, we recall a calculation presented in \citep{bib:bratek_MNRAS}, which implies that in thin disk model the relative contribution to the local surface density from an almost Keplerian tail can be neglected at radii lower
than roughly $2/3$ of the radial extent of the measured part of the rotation curve, provided the mass function of the flattened mass component can be assumed to be almost saturated at the last measurement point. Then, in practice, it would make no difference if we set the rotation to be zero beyond the last measured point or not.

\begin{figure}
   \centering
      \includegraphics[width=0.5\textwidth
     ,trim=16pt 18pt 4pt 16pt
      ,clip=true
      ]{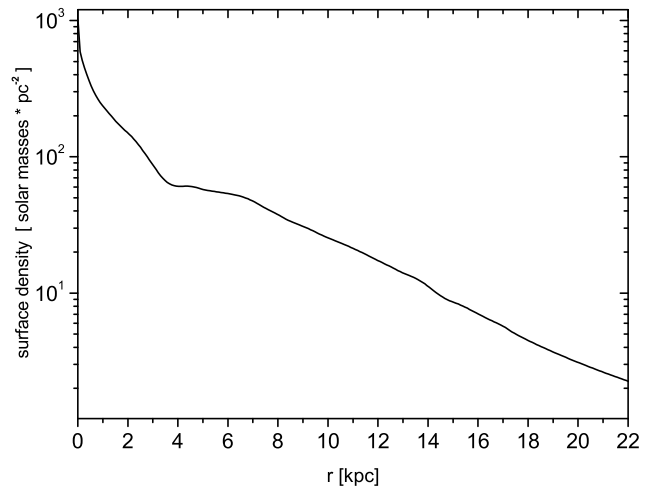}
      \caption{\label{fig:sig} Surface mass density accounting in thin disk model for the adopted rotation curve of galaxy \ngc shown in \figref{fig:rot}. Unlike in spherical symmetry, surface density is nonzero in a region $r>17\,\kpc$ with Keplerian rotation curve.
}
  \end{figure}

The surface density corresponding to the adopted rotation curve shown in  \figref{fig:rot} is presented in \figref{fig:sig}.
It should be stressed that in disk model, unlike in spherical symmetry, the
Keplerian tail in a region is not equivalent to lack of matter in
that region, which is readily seen in \figref{fig:sig}, where surface density is indeed non-vanishing in the region beyond $\approx17\,\kpc$, where the rotation curve was set to be Keplerian.

\section{Vertical gradient in azimuthal velocity}

In the quasi-circular orbits approximation \citep{bib:gradient},
the azimuthal velocity at an altitude $z$ above the mid-plane is
found from
\begin{equation}\label{eq:veloc}
{v_{\varphi}^2(r,z)}=\!\int\limits_0^{\infty}\!\!\!\frac{2\, G\,\sigma(\chi)\,\chi\,\ud{\chi{}}}{\sqrt{\br{r+\chi{}}^2+z^2}}\!\br{\!K\sq{\kappa}-
\frac{\chi^2-r^2+z^2}{\br{r-\chi}^2+z^2}\,E\sq{\kappa}\!}\!,\end{equation}
\begin{equation*} \kappa=\sqrt{\frac{4r\chi{}}{\br{r+\chi{}}^2+z^2}},\end{equation*}
where the surface density expressed by \eqref{eq:SigmafromRotCrv} have been substituted for $\sigma(\chi)$.
This relation is used to describe the behavior of the
azimuthal velocity component in a region above the mid-plane and exterior to the main concentration of masses of the flattened mass distribution, where the vertical structure of this distribution can be neglected, but still close enough to the mid-plane, so that the projection of orbits onto the mid-plane could be regarded as a perturbation of the concentric circular orbits based on which \eqref{eq:SigmafromRotCrv} turning the circular velocity to the surface density (and its inverse) are derived in thin disk model (though, a reasoning in \citep{bib:gradient} admits a more general class of orbits).

\medskip

\begin{figure}
   \centering
      \includegraphics[width=0.5\textwidth
      ,trim=16pt 18pt 4pt 16pt
      ,clip=true
      ]{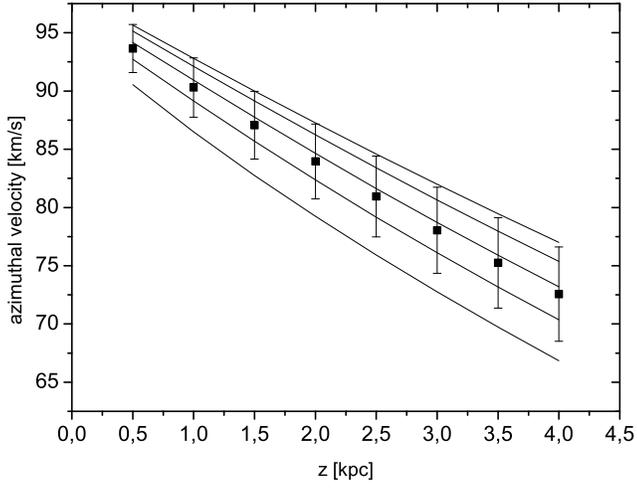}
      \caption{\label{fig:veldecr} Behavior of azimuthal velocity off the mid-plane in galaxy \ngc, predicted in quasi-circular orbits approximation in thin disk model (that is, calculated with the help of integral \eqref{eq:veloc}).
The thin lines are curves of constant radial variable: $r=8,9,10,11,12\,\kpc$, counting from  the bottom line to the top line.
Each
point represents the mean azimuthal velocity component, with the respective bar representing the standard deviation. The velocity decreases almost linearly with the altitude $\abs{z}$ above the mid-plane.
}
  \end{figure}

We will estimate the vertical gradient in azimuthal
velocity with the help of  two methods.
The first method gives a mean magnitude of the gradient by means of a linear approximation of
the azimuthal velocity.
The mean value, denoted by $\gamma$,  is determined in a given rectangular region
$(r,z)$ by finding a best fitting profile of the form
\begin{equation}\label{eq:model_rot}v_{\varphi}(r,z)=v_{\varphi}(r,0)+\gamma
\,\abs{z}+\delta\, z,\end{equation}
approximating the values of azimuthal velocity
stipulated at several points arranged in an array
 in that region. The linear
formula was used by \citet{bib:gradmlecz} to estimate the vertical
fall-off rate in the rotation speed of the Milky-Way galaxy.
Here, the use of the linear expansion is justified by the linearity of the decrease in azimuthal velocity predicted by the thin disk model for \ngc, as illustrated in \figref{fig:veldecr}.
Owing to this behavior (and the reflection symmetry with respect to the mid-plane $z=0$, which implies that the rolling parameter $\delta$ must be set equal to zero in disk model), it is rational to consider only a single number describing the vertical velocity fall-off for \ngc, namely, the mean gradient magnitude $\gamma$.

Table \ref{tab:table} shows the results of applying the first
method
 to the theoretical rotation values predicted by evaluating the integral \eqref{eq:veloc}
 inside several
 rectangular regions defined in that table.
\begin{table}\centering
\begin{tabular}{l@{\xspace\xspace}cl@{\xspace\xspace}c|c}
\hline
& $\delta r$ & & $\delta z$  &\ \  $\frac{\partial{v_{\phi}}}{\partial{z}}$ $[\,\grad ]$\\
\hline
$r\!\in\!(1 ,16 ) $ & $   0.2 $ &
$z\!\in\!(0.5 , 1 )$ & $   0,02  $
&   $-8.38\pm 0.33 $\\
$r\!\in\!(2 , 10 ) $ & $ 0.2 $&
$z\!\in\!(0.5 , 3 ) $ & $   0.2 $&
$-8.24\pm 0.21 $\\
$r\!\in\!(10 , 16 )$ & $   0.2 $&
$z\!\in\!(0.5 , 1 ) $ & $   0.02 $&
$-5.23\pm 0.05 $\\
$r\!\in\!(10 , 11 ) $ & $   0.01 $&
$z\!\in\!(0.5 , 3 )$ & $   0.2 $&
$-6.00\pm 0.00 $\\ 
\hline
\end{tabular}
\caption{\label{tab:table}Rectangular arrays of rotation data ($z$, $r$ , $\delta z$ and  $\delta r$, are expressed in $\,\kpc$) and
the corresponding mean vertical gradient of rotation (coefficient $\gamma$ in \eqref{eq:model_rot}).}
\end{table}
To some degree, the gradient results depend on the horizontal and vertical extent of a given region where the gradient is estimated.
To trace this dependence we use a second  method.

\medskip

The second method   of determining the
vertical gradient in azimuthal velocity uses a
direct expression for the gradient, obtained by differentiating
 the expression for velocity \eqref{eq:veloc} with respect to $z$
and using some identities between elliptic integrals $K,E$ and their
derivatives.  The result is
\begin{eqnarray}\label{eq:diskgradient}\partial_z{}v_{\varphi}\br{r,z}=\frac{Gz}{v_{\varphi}(r,z)}\int\limits_0^{\infty} \frac{\chi \sigma (\chi )\ud{\chi}}{{\left( z^2 + {\left( r + \chi  \right) }^2 \right) }^{\frac{3}{2}}}\times \end{eqnarray}
   \[
\left( \frac{r^2 - z^2 - {\chi }^2}{z^2 + {\left( r - \chi  \right) }^2}K(\kappa)       -
\!\frac{7r^4 \!+\! 6r^2\!\left(\! z^2\! -\! {\chi }^2 \right)\!  -\! {\left(
\! z^2\! +\! {\chi }^2 \right)\! }^2}
  {{\left( z^2 + {\left( r - \chi  \right) }^2 \right) }^2}E(\kappa)
  \right).\]

\begin{figure}
   \centering
      \includegraphics[width=0.5\textwidth
      ,trim=16pt 18pt 4pt 16pt
      ,clip=true
      ]{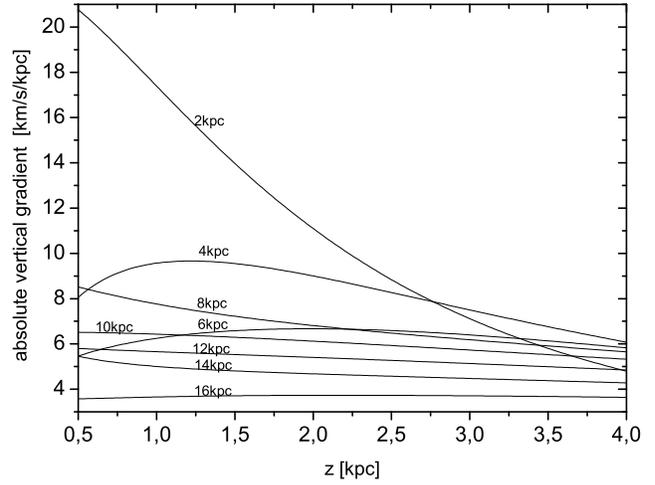}
      \caption{\label{fig:grad1}
Absolute values of the vertical gradient in azimuthal velocity in thin disk model shown in function of the
altitude $z$ above the mid-plane for various values of the radial
variable from within the interval $r\in\br{2,16}\,\kpc$ in steps of $2 \,\kpc$. These lines represent the values obtained using the integral expression \eqref{eq:diskgradient}.}
  \end{figure}

\noindent
\figref{fig:grad1}
illustrates the gradient dependence in function of the altitude above the mid-plane for various radii in a region from $2\,\kpc$ to $16\,\kpc$.
The absolute gradient value depends weakly on the $z$ variable, except close to the galactic center. This reflects the linear decrease of azimuthal velocity evident in \figref{fig:veldecr}.
The maximum gradient magnitude exceeds even $20\,\grad $ but only close to the center and at low heights above the mid-plane.
In a prevailing part of the galactic region the gradient magnitude does not exceed  $10\,\grad $, decreasing moderately with the radial distance and is lower than $4\,\grad $ for the outermost radii (compare, the $r=16\,\kpc $ line in \figref{fig:grad1}).

\begin{figure}
   \centering
      \includegraphics[width=0.5\textwidth
      ,trim=16pt 18pt 4pt 16pt
      ,clip=true
      ]{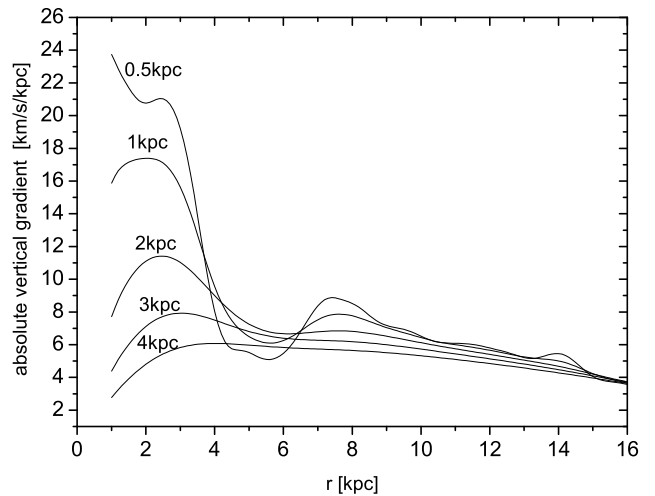}
      \caption{\label{fig:grad2} Absolute values of the vertical gradient in azimuthal velocity shown as functions of the radial variable for various altitudes above the mid-plane.}
  \end{figure}

\figref{fig:grad2}
shows for several altitudes the variation in the gradient in function of the radial distance. In the central part the absolute gradient value is not only high (exceeding $20\,\grad $ for low altitudes) but it also strongly changes with the radial variable. But starting from a radius of $4 \,\kpc$ the gradient magnitudes are much lower (below $10\,\grad $) and they exhibit a significantly weaker dependence on the radial variable, being more pronounced only for lower altitudes above the mid-plane.

\medskip

We can summarize the above results by
saying that the vertical gradient predicted for \ngc is not too high -- its mean value in a rectangular region $r\in(1,16)\,\kpc$ and $\abs{z}\in(0.5,1)\,\kpc$ is $-8.38\pm 0.33\,\grad $, and it would not differ much
from a value of $-8.24\pm0.21\,\grad$ obtained by shrinking the radial interval to $r\in(2, 10)\,\kpc$ and increasing the height above the mid-plane to $3\,\kpc$.
We predict also a
decrease in the gradient magnitude for higher radial distances: in a rectangle
$r\in(2,16)\,\kpc$ and $\abs{z}\in(0.5,1)\,\kpc$
we obtain a mean gradient value of $-5.23\pm0.05\,\grad $.
These predictions are in agreement to within error limits with the gradient properties reported in \citep{bib:ngc4244}, namely, with the gradient value of $-9^{+3}_{-2}\,\grad $ in the approaching half and $-9^{+2}_{-2} \,\grad $ in the receding half, and with
 the gradient magnitude found to be decreasing to $5^{+2}_{-2}\,\grad $ and $4^{+2}_{-2} \,\grad $ near a radius of $10 \,\kpc$ in the
approaching and receding halves, respectively.

\begin{figure}
   \centering
      \includegraphics[width=0.5\textwidth
      ,trim=16pt 18pt 4pt 16pt
      ,clip=true
      ]{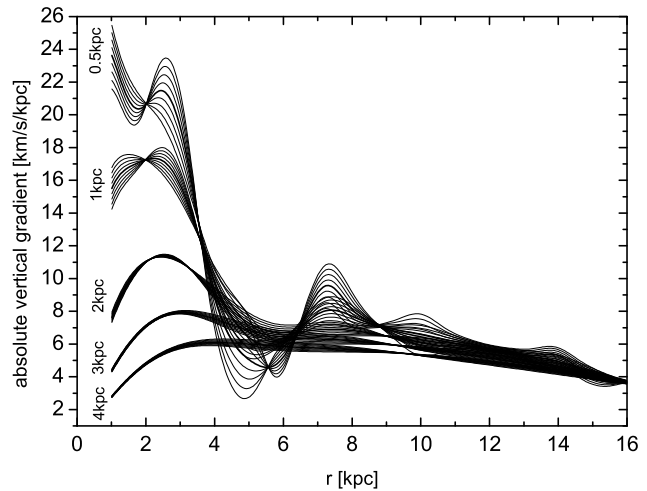}
      \caption{\label{fig:graderr} This figure shows results analogous to those in \figref{fig:grad2} for various rotation curves. Thick lines correspond to a rotation curve obtained by smoothing spline interpolating between the approaching- and receding- half rotation curves (it almost overlaps the adopted rotation curve in \figref{fig:rot}, which shows that the averaging method is not influential to the overall shape of that curve).
Thin lines correspond to deformed rotation curves obtained by taking linear combinations $\alpha v_a+(1-\alpha)v_r$ of the rotation curve in the  approaching half (denoted here by $v_a$)  and that in the receding half (denoted by $v_r$). }
  \end{figure}

In the first method we defined the error of our model gradient value as  the uncertainty in the parameter $\gamma$ inferred from a linear regression model \eqref{eq:model_rot} fitting to the velocity values predicted for quasi-circular orbits. The error does not include the error transferred from the uncertainties in the rotation curve determination.
We can get a good impression of the scale of these uncertainties from  \figref{fig:graderr} which illustrates the influence of variations in the rotation curve on the predicted gradient values.
But the relative errors in the gradient determination
in \citep{bib:ngc4244} are even higher, therefore a more precise analysis of the model uncertainties seems of minor importance for our conclusions concerning the gradient properties.

\section{Concluding remarks}

There are various models aimed at explaining the presence and behavior of the vertical gradient of azimuthal velocity (e.g., \citet{bib:collins,bib:frabi}),
but our model is
characterized by high simplicity and minimum of assumptions.
We use a disk model with a surface density accounting for the full rotation curve measured for a galaxy, assuming that the motion
of matter is entirely due to the gravitational potential of the
disk.
This means that most of matter is assumed to form a flattened
mass distribution, without a dominating heavy halo.
Despite this simplicity our model accounts for the vertical gradient in azimuthal velocity, predicting correct gradient magnitudes and its approximate independence of the altitude above the mid-plane.

 \ngc is the next among the galaxies examined by us so far (\citet{bib:gradient,bib:gradientwie}) for which our approach to modeling the gradient turned out  to perform satisfactorily well, predicting vertical gradient magnitudes and their behavior consistently with expectations based on the measurements. The most controversial in this approach
may be our not including the other mass components except the disk-like only.
But obtaining satisfactory results in modeling a galaxy in this approach, simply may indicate that the galaxy has a dominating flattened mass component, at least in the region where the gradient is measured and modeled. This we give as the main conclusion of our work.

An interesting property we want to bring to the attention are high gradient magnitudes in the galactic center vicinity of \ngc
(mainly close to the mid-plane), exceeding the mean gradient magnitude determined for this galaxy. A similar phenomenon occurs for another galaxy.
\citet{bib:ngc4244} mention that  the measured gradient magnitude in the central part of NGC 891 is very high (reaching $43\,\grad $) although it decreases strongly for outer radii (to $14\,\grad $). This may suggest that high gradient magnitudes in the central parts of spiral galaxies (much higher than the mean magnitude for the whole disk) might be a qualitative property of the vertical gradient which the disk model accounts for well.

\bibliography{grad4244}
\bibliographystyle{mn2e}

\end{document}